\documentclass[12pt]{article}
\usepackage{amsmath}
\usepackage{graphicx}
\usepackage{color}
\begin{document}
\baselineskip=18 pt
\begin{center}
{\large{\bf Effects of linear Central potential induced by Lorentz symmetry breaking on a generalized Klein-Gordon Oscillator }}
\end{center}

\vspace{0.5cm}

\begin{center}
{\bf Faizuddin Ahmed}\footnote{\bf faizuddinahmed15@gmail.com ; faiz4U.enter@rediffmail.com}\\
{\bf National Academy Gauripur, Assam, 783331, India }
\end{center}

\vspace{0.5cm}

\begin{abstract}

We investigate the generalized Klein-Gordon oscillator under the Lorentz symmetry breaking effects where, a linear electric and constant magnetic field is considered and analyze its effects on the relativistic quantum oscillator. Furthermore, the behaviour of the quantum oscillator in the presence of a Cornell-type scalar potential is analyzed and the solution of the bound state is obtained. We see that the analytical solution to the generalized Klein-Gordon oscillator can be achieved and the angular frequency of the oscillator depends on the quantum numbers of the system.

\end{abstract}

{\bf Keywords}: Lorentz symmetry violation, Relativistic wave-equations, scalar potential, electric \& magnetic field, biconfluent Heun Equation.

\vspace{0.3cm}

{\bf PACS Number(s):} 03.65.Pm,  11.30.Cp, 11.30.Qc

\section{Introduction}

Inspired by the Dirac oscillator model for spin $\frac{1}{2}$-fermionic field \cite{ff1,ff2}, a relativistic oscillator model for spin-$0$ scalar field was proposed in Ref. \cite{ff3,BM} which is known as the Klein–Gordon oscillator. For the Klein-Gordon oscillator, one can replace the momentum operator by $\vec{p} \rightarrow \vec{p}-i\,M\,\omega\,r\,\vec{r}$ in the Klein-Gordon equation. Here $\omega$ is the angular frequency of the oscillator, and $r$ being the axial/radial distance. The Klein–Gordon oscillator has been studied in the quantum mechanics, such as, In non-commutative phase space with a magnetic field \cite{ff12}, in a space-time with cosmic string \cite{AB}, in a G\"{o}del-type space-time \cite{ZW}, effects of linear and Coulomb-type central potentials \cite{ff4,ff5,ff6}, in Kaluza-Klein theory \cite{JC}, in a space-time with screw dislocation under a scalar potential \cite{ff10}, in a space-time with torsion \cite{ff7}, with non-inertial effects \cite{ff8}, with position-dependent mass in rotating cosmic string space-time \cite{ff13}, in cosmic string space-time under a Cornell-type potential \cite{ff9}, in $(1+2)$-dimensional space-time \cite{AOP}, in a cosmic string space-time with spacelike dislocation under Cornell-type potential \cite{AHEP}, effects of global monopole space-time \cite{ff11}.

In this paper, we study the generalized Klein-Gordon oscillator \cite{a28,a35,a29,a30,a34,a37,a33} under  Lorentz symmetry breaking defined by a tensor $(K_F)_{\mu\nu\alpha\beta}$ that governs the Lorentz symmetry violation out of the Standard Model Extension \cite{a3,a4,a6,a7,a77,a777,a8,a88,a888}. The violation of the Lorentz symmetry is determined by a non-null component of tensor, a linear electric and constant magnetic field. Then, we search for the solution of the bound state of the generalized Klein-Gordon oscillator in the presence of a scalar potential by modifying the mass term. We see that the angular frequency of the oscillator depends on the quantum numbers of the system.

The Standard Model Extension (SME) \cite{a3,a4,a6,a7,a77,a777,a8,a88,a888} is an extension of the Standard Model of the fundamental interactions where, the effective Lagrangian corresponds to the usual Standard Model plus Lorentz-violating (LV) tensorial background coefficients. The gauge sector of SME consists of two terms called CPT-odd sector \cite{a3,a4} and CPT-even sector \cite{b1,a5}. The relativistic quantum systems under Lorentz symmetry violation has been studied in Refs. \cite{a5,b1,b3,b4,b5,b6,b10,b11,b12}. The Lorentz symmetry violation has investigated at a low energy scenario in Refs. \cite{b13,b14,b15}.

The Klein-Gordon equation under Lorentz symmetry violation background is given by \cite{b6,b10,b11,b12}
\begin{equation}
p^{\mu}\,p_{\mu} \rightarrow p^{\mu}\,p_{\mu}+\frac{\alpha}{4}\,(K_F)_{\mu\nu\alpha\beta}\,F^{\mu\nu} (x)\,F^{\alpha\beta}(x),
\label{aa}
\end{equation}
where $\alpha$ is a constant, $F_{\mu\nu}(x)$ is the electromagnetic field tensor, $(K_F)_{\mu\nu\alpha\beta}$ is a dimensionless tensor. It has the symmetries of the Riemann tensor $R_{\mu\nu\alpha\beta}$ and zero on double trace, so it contains 19 independent real components. 

Based on the coupling of the generalized Dirac-oscillator as done in Ref. \cite{KB}, a generalized oscillator model to the Klein-Gordon field was given in Refs. \cite{a28,a35,a29,a30,a34,a37,a33} where the momentum operator is changed as
\begin{equation}
p_{\mu} \rightarrow p_{\mu}-i\,M\,\omega\,X_{\mu}
\label{osc}
\end{equation}
where $\omega$ is the oscillator frequency, $X_{\mu}=(0, f(r), 0, 0)$ is the four-vector and $f (r)$ is an arbitrary function.

\section{Generalized KG-oscillator Under the Effects of Lorentz Symmetry Violation }

The generalized Klein-Gordon oscillator under the effects of Lorentz symmetry breaking (\ref{aa}) becomes
\begin{equation}
p^{\mu}\,p_{\mu} \rightarrow (p^{\mu}+i\,M\,\omega\,X^{\mu})\,(p_{\mu}-i\,M\,\omega\,X_{\mu})+\frac{\alpha}{4}\,(K_F)_{\mu\nu\alpha\beta}\,F^{\mu\nu} (x)\,F^{\alpha\beta}(x).
\label{aa2}
\end{equation}

Therefore, the generalized Klein-Gordon oscillator equation is given by ($c=\hbar=1$):
\begin{equation}
\left[(p^{\mu}+i\,M\,\omega\,X^{\mu})\,(p_{\mu}-i\,M\,\omega\,X_{\mu})+\frac{\alpha}{4}\,(K_F)_{\mu\nu\alpha\beta}\,F^{\mu\nu} (x)\,F^{\alpha\beta} (x)-M^2\right]\,\Psi=0.
\label{2}
\end{equation}

We consider the Minkowski flat space-time in the cylindrical coordinates $(t, r, \phi, z)$, where coordinates have their usual ranges given by
\begin{equation}
ds^2=-dt^2+dr^2+r^2\,d\phi^2+dz^2.
\label{3}
\end{equation}
Therefore, the generalized Klein-Gordon oscillator under the effects of the Lorentz symmetry violation equation (\ref{2}) becomes
\begin{eqnarray}
&&\left[-\frac{\partial^2}{\partial t^2}+\frac{1}{r}\,\left (\frac{\partial}{\partial r}+M\,\omega\,f(r) \right)\,\left (r\,\frac{\partial}{\partial r}-M\,\omega\,r\,f(r) \right)+\frac{\partial^2}{\partial z^2}+\frac{1}{r^2}\,\frac{\partial^2}{\partial \phi^2} \right]\,\Psi\nonumber\\
&&+\frac{\alpha}{4}\,(K_F)_{\mu\nu\alpha\beta}\,F^{\mu\nu} (x)\,F^{\alpha\beta}(x)\,\Psi=M^2\,\Psi.
\label{4}
\end{eqnarray}

Using the properties of the tensor $(K_F)_{\mu\nu\alpha\beta}$ given in Refs. \cite{a3,a4,a6,a7,a77,a777,a8,a88,a888,a5,b1,b3,b4,b5,b6,b10,b11,b12}, we can rewrite equation (\ref{4}) as
\begin{eqnarray}
&&\left[-\frac{\partial^2}{\partial t^2}+\frac{\partial^2}{\partial r^2}+\frac{1}{r}\,\frac{\partial}{\partial r}+\frac{1}{r^2}\,\frac{\partial^2}{\partial \phi^2}+\frac{\partial^2}{\partial z^2}-M^2\,\omega^2\,f^2 (r)-M\,\omega\,\left(f' (r)+\frac{f (r)}{r} \right) \right]\,\Psi\nonumber\\
&&+\left[-\frac{\alpha}{2}\,(\kappa_{DE})_{ij}\,E^{i}\,E^{j}+\frac{\alpha}{2}\,(\kappa_{HB})_{jk}\,B^i\,B^j-\alpha\,(\kappa_{DB})_{jk}\,E^i\,B^j \right]\,\Psi\nonumber\\
&&=M^2\,\Psi.
\label{5}
\end{eqnarray}

Let us consider a possible scenario of the Lorentz symmetry violation determined by only one non-null component of the tensor $(\kappa_{DB})_{jk}$ as being $(\kappa_{DB})_{13}=\kappa= const$ and by a field configuration given by \cite{b6,b10}:
\begin{equation}
\vec{B}=B_0\,\hat{z}\quad,\quad \vec{E}=\frac{\lambda}{2}\,r\,\hat{r}
\label{6}
\end{equation}
where $B_0$ is a constant, $\hat{z}$ is a unit vector in the $z$-direction and $\lambda$ is a constant associated with linear charge density of electric charge along the axial direction.

Hence Equation (\ref{5}) becomes
\begin{eqnarray}
&&\left[-\frac{\partial^2}{\partial t^2}+\frac{\partial^2}{\partial r^2}+\frac{1}{r}\,\frac{\partial}{\partial r}+\frac{1}{r^2}\,\frac{\partial^2}{\partial \phi^2}+\frac{\partial^2}{\partial z^2}-M^2\,\omega^2\,f^2 (r)-M\,\omega\,\left(f' (r)+\frac{f (r)}{r} \right) \right]\,\Psi\nonumber\\
&&-\frac{\alpha\,\lambda\,B_0\,\kappa}{2}\,r\,\Psi=M^2\,\Psi.
\label{7}
\end{eqnarray}

Let the solution to the Eq. (\ref{7}) is
\begin{equation}
\Psi (t, r, \phi, z)=e^{i\,(-E\,t+l\,\phi+k\,z)}\,\psi (r),
\label{8}
\end{equation}
where $E, l, k$ have their usual meaning.

Substituting the solution (\ref{8}) into the Eq. (\ref{7}), we obtain the following radial wave-equation for $\psi(r)$:
\begin{eqnarray}
&&\psi''(r)+\frac{1}{r}\,\psi' (r)+\left[E^2-k^2-\frac{l^2}{r^2}-M^2\,\omega^2\,f^2-M\,\omega\,\left(f'+\frac{f }{r} \right)\right]\,\psi (r)\nonumber\\
&&-\frac{\alpha\,\lambda\,B_0\,\kappa}{2}\,r\,\psi (r)=M^2\,\psi (r).
\label{9}
\end{eqnarray}

To study the generalized KG-oscillator under the effects of Lorentz symmetry violation, we have chosen the following function \cite{KB}
\begin{equation}
f(r)=b_1\,r+\frac{b_2}{r},
\label{10}
\end{equation}
where $b_1 > 0, b_2 > 0$ are arbitrary constants. This type of function has been studied in the relativistic quantum systems Refs. \cite{a31,a32,a36,a28,a35,a29,a30,a34,a33,a37}.

Substituting the function (\ref{10}) into the Eq. (\ref{9}), we obtain the following radial wave-equation equation :
\begin{equation}
\left [\frac{d^2}{dr^2}+\frac{1}{r}\,\frac{d}{dr}+\Lambda-M^2\,\omega^2\,b^2_{1}\,r^2-\frac{j^2}{r^2}-b\,r \right]\,\psi (r)=0,
\label{11}
\end{equation}
where
\begin{eqnarray}
&&\Lambda=E^2-M^2-k^2-2\,M\,\omega\,b_1-2\,M^2\,\omega^2\,b_1\,b_2,\nonumber\\
&&j=\sqrt{l^2+M^2\,\omega^2\,b^2_{2}},\nonumber\\
&&b=\frac{\alpha\,\lambda\,B_0\,\kappa}{2}.
\label{12}
\end{eqnarray}

Transforming $x=\sqrt{M\,\omega\,b_1}\,r$ into the Eq. (\ref{14}), we obtain the following equation:
\begin{equation}
\left [\frac{d^2}{dx^2}+\frac{1}{x}\,\frac{d}{dx}+\zeta-x^2-\frac{j^2}{x^2}-\theta\,x \right]\,\psi (x)=0,
\label{13}
\end{equation}
where
\begin{equation}
\zeta=\frac{\Lambda}{M\,\omega\,b_1}\quad,\quad \theta=\frac{b}{(M\,\omega\,b_1)^{\frac{3}{2}}}.
\label{14}
\end{equation}

Now, we use the appropriate boundary conditions that the wave functions is regular both at $x \rightarrow 0$ and $x \rightarrow \infty $. Suppose the possible solution to the Eq. (\ref{13}) is
\begin{equation}
\psi (x)=x^{j}\,e^{-\frac{1}{2}\,(x+\theta)\,x}\,H (x).
\label{15}
\end{equation}
Substituting the solution (\ref{15}) into the Eq. (\ref{13}), we obtain the following equation
\begin{equation}
H'' (x)+\left[\frac{1+2\,j}{x}-2\,x-\theta \right]\,H' (x)+\left[-\frac{\frac{\theta}{2}\,(1+2\,j)}{x}+\Theta \right]\,H (x)=0,
\label{16}
\end{equation}
where $\Theta=\zeta+\frac{\theta^2}{4}-2\,(1+j)$. 

Equation (\ref{16}) is the biconfluent Heun's differential equation \cite{ERFM,a29,a30,a33,a34,AR,SYS} with $H(x)$ is the Heun polynomials function.

The above equation (\ref{16}) can be solved by the Frobenius method. Writing the solution as a power series expansion around the origin \cite{GBA}:
\begin{equation}
H (x)=\sum^{\infty}_{i=0}\,d_{i}\,x^{i}.
\label{18}
\end{equation}
Substituting the power series solution into the Eq. (\ref{16}), we obtain the following recurrence relation
\begin{equation}
d_{n+2}=\frac{1}{(n+2)(n+2+2\,j)}\,\left[\theta\,(n+j+\frac{3}{2})\,d_{n+1}-(\Theta-2\,n)\,d_n \right].
\label{19}
\end{equation}
With few coefficients are
\begin{eqnarray}
&&d_1=\frac{\theta}{2}\,d_0,\nonumber\\
&&d_2=\frac{1}{4\,(1+j)}\,\left[\theta\,(j+\frac{3}{2})\,d_1-\Theta\,d_0 \right].
\label{20}
\end{eqnarray}

The power series expansion $H(x)$ becomes a polynomial of degree $n$ by imposing the following two conditions \cite{ERFM,a29,a30,a33,a34}
\begin{eqnarray}
\Theta&=&2\,n,\quad (n=1,2,...)\nonumber\\
d_{n+1}&=&0.
\label{21}
\end{eqnarray}

By analyzing the first condition, we obtain following equation for the energy eigenvalue $E_{n,l}$:
\begin{equation}
E_{n,l}=\pm\,\sqrt{M^2+k^2+2\,M^2\,\omega^2\,b_1\,b_2+2\,M\,\omega\,b_1\left(n+2+\sqrt{l^2+M^2\,\omega^2\,b^2_{2}}\right)-\left(\frac{\alpha\,\lambda\,B_0\,\kappa}{4\,M\,\omega\,b_1}\right)^2}.
\label{22}
\end{equation}
Note that Eq. (\ref{22}) is not the general expression of the relativistic energy eigenvalues of the generalized KG-oscillator field. 
\begin{equation}
E_{n,l}=\pm\,\sqrt{M^2+k^2+2\,M\,\omega\,b_1\left(n+2+|l|\right)-\left(\frac{\alpha\,\lambda\,B_0\,\kappa}{4\,M\,\omega\,b_1}\right)^2}.
\label{220}
\end{equation}
which is similar to the result obtained in Ref. \cite{b10} (see Eq. (19) in the Ref. \cite{b10}). 

The radial wave-functions are given by
\begin{equation}
\psi_{n,l} (x)=x^{\sqrt{l^2+M^2\,\omega^2\,b^2_{2}}}\,e^{-\frac{1}{2}\,\left[x+\frac{\alpha\,\lambda\,B_0\,\kappa}{2\,(M\,\omega\,b_1)^{\frac{3}{2}}}\right]\,x}\,H (x).
\label{23}
\end{equation}

Now, we evaluate the individual energy level and eigenfunction one by one as in Refs. \cite{ERFM,a29,a30,a33,a34}. For example, $n=1$, we have $\Theta=2$ and $d_2=0$ which implies
\begin{eqnarray}
&&\Rightarrow \frac{2}{\theta\,(j+\frac{3}{2})}\,d_0=\frac{\theta}{2}\,d_0\nonumber\\ 
&&\Rightarrow \omega_{1,l}=\left [\frac{(\alpha\,\lambda\,B_0\,\kappa)^{\frac{2}{3}}\,(j+\frac{3}{2})^{\frac{1}{3}}}{4\,M\,b_1} \right]
\label{24}
\end{eqnarray}
a constraint on the angular frequency $\omega_{1,l}$ of the oscillator. Note that its value changes for each quantum number $n$ and $l$ of the system, so we have labeled $\omega \rightarrow \omega_{n,l}$. 

Therefore, the ground state energy level for the radial mode $n=1$ is given by
\begin{equation}
E_{1,l}=\pm\,\sqrt{M^2+k^2+2\,M\,\omega_{1,l}\,b_1\,\left(M\,\omega_{1,l}\,b_2+3+\sqrt{l^2+M^2\,\omega^2_{1,l}\,b^2_{2}}\right)-\left(\frac{\alpha\,\lambda\,B_0\,\kappa}{\frac{3}{2}+\sqrt{l^2+M^2\,\omega^2_{1,l}\,b^2_{2}}}\right)^{\frac{2}{3}}}.
\label{26}
\end{equation}
And the ground state eigenfunction is 
\begin{equation}
\psi_{1,l} (x)=x^{\sqrt{l^2+M^2\,\omega^2\,b^2_{2}}}\,e^{-\frac{1}{2}\,(x+2\,d_1)\,x}\,(1+d_1\,x).
\label{27}
\end{equation}
where for simplicity $d_0=1$ and
\begin{equation}
d_1=\frac{2}{\left(\sqrt{l^2+M^2\,\omega^2_{1,l}\,b^2_{2}}+\frac{3}{2}\right)^{\frac{1}{2}}}.
\label{28}
\end{equation}
The lowest energy state (\ref{26}) plus the ground state wave-function (\ref{27})--(\ref{28}) with the restriction (\ref{24}) on the angular frequency of the oscillator is defined for the radial mode $n=1$. 

\section{Generalized KG-oscillator Under the Effects of Lorentz Symmetry Violation subject to Cornell-type potential }

We introduce a scalar potential $S(r)$ in the generalized KG-oscillator by modifying the mass term $M \rightarrow M+ S (r)$ \cite{ERFM} under the effects of the Lorentz symmetry violation. Therefore, the radial wave-equation for $\psi(r)$ the Eq. (\ref{9}) becomes
\begin{eqnarray}
&&\psi''(r)+\frac{1}{r}\,\psi' (r)+\left[E^2-k^2-\frac{l^2}{r^2}-M^2\,\omega^2\,f^2-M\,\omega\,\left(f'+\frac{f }{r} \right)\right]\,\psi (r)\nonumber\\
&&-\frac{\alpha\,\lambda\,B_0\,\kappa}{4}\,r\,\psi (r)=(M+S(r))^2\,\psi (r).
\label{29}
\end{eqnarray}

In this work, we have chosen a Cornell-type scalar potential given by \cite{a29,a30,ERFM,EE,EE2,EE3,b8,b9}
\begin{equation}
S (r)=\eta_L\,r+\frac{\eta_c}{r},
\label{30}
\end{equation}
where $\eta_L>0, \eta_c>0$ are arbitrary constants. 

Substituting the above potential (\ref{30}) and the function (\ref{10}) into the radial wave-equation (\ref{29}), we have
\begin{equation}
\left [\frac{d^2}{dr^2}+\frac{1}{r}\,\frac{d}{dr}+\tilde{\Lambda}-\Omega^2\,r^2-\frac{\tilde{j}^2}{r^2}-\frac{a}{r}-\tilde{b}\,r \right]\,\psi (r)=0,
\label{31}
\end{equation}
where we have defined
\begin{eqnarray}
&&\tilde{\Lambda}=E^2-M^2-k^2-2\,M\,\omega\,b_1-2\,M^2\,\omega^2\,b_1\,b_2-2\,\eta_L\,\eta_c,\nonumber\\
&&\Omega=\sqrt{M^2\,\omega^2\,b^2_{1}+\eta^2_{L}},\nonumber\\
&&\tilde{j}=\sqrt{j^2+\eta^2_{c}},\nonumber\\
&&a=2\,M\,\eta_c,\nonumber\\
&&\tilde{b}=b+2\,M\,\eta_L.
\label{32}
\end{eqnarray}

Transforming $x=\sqrt{\Omega}\,r$ in the above equation (\ref{31}), we have
\begin{equation}
\left [\frac{d^2}{dx^2}+\frac{1}{x}\,\frac{d}{dx}+\tilde{\zeta}-x^2-\frac{\tilde{j}^2}{x^2}-\frac{\eta}{x}-\tilde{\theta}\,x \right]\,\psi (x)=0,
\label{33}
\end{equation}
where
\begin{equation}
\tilde{\zeta}=\frac{\tilde{\Lambda}}{\Omega}\quad,\quad \eta=\frac{a}{\sqrt{\Omega}}\quad,\quad \tilde{\theta}=\frac{\tilde{b}}{\Omega^{\frac{3}{2}}}.
\label{34}
\end{equation}

Suppose the possible solution to the Eq. (\ref{33}) is
\begin{equation}
\psi (x)=x^{\tilde{j}}\,e^{-\frac{1}{2}\,(x+\tilde{\theta})\,x}\,H (x).
\label{35}
\end{equation}
Substituting the solution (\ref{35}) into the Eq. (\ref{33}), we obtain the following equation
\begin{equation}
H'' (x)+\left[\frac{1+2\,\tilde{j}}{x}-2\,x-\tilde{\theta} \right]\,H' (x)+\left[-\frac{\beta}{x}+\tilde{\Theta} \right]\,H (x)=0,
\label{36}
\end{equation}
where
\begin{equation}
\tilde{\Theta}=\tilde{\zeta}+\frac{\tilde{\theta}^2}{4}-2\,(1+\tilde{j})\quad,\quad \beta=\eta+\frac{\tilde{\theta}}{2}\,(1+2\,\tilde{j}).
\label{37}
\end{equation}
Equation (\ref{36}) is the biconfluent Heun's differential equation \cite{ERFM,a29,a30,a33,a34,AR,SYS} with $H(x)$ is the Heun polynomials function.

Considering the power series method considered earlier, we obtain the following recurrence relations
\begin{equation}
d_{n+2}=\frac{1}{(n+2)(n+2+2\,\tilde{j})}\,\left[\left\{\beta+\tilde{\theta}\,(n+1) \right\}\,d_{n+1}-(\tilde{\Theta}-2\,n)\,d_n \right].
\label{38}
\end{equation}
With few coefficients are
\begin{eqnarray}
&&d_1=\left(\frac{\eta}{1+2\,\tilde{j}}+\frac{\tilde{\theta}}{2} \right)\,d_0,\nonumber\\
&&d_2=\frac{1}{4\,(1+\tilde{j})}\,[(\beta+\tilde{\theta})\,d_1-\tilde{\Theta}\,d_0 ].
\label{39}
\end{eqnarray}

The power series expansion $H(x)$ becomes a polynomial of degree $n$ by imposing the following two conditions \cite{ERFM,a29,a30,a33,a34}
\begin{eqnarray}
\tilde{\Theta}&=&2\,n,\quad (n=1,2,...)\nonumber\\
d_{n+1}&=&0.
\label{40}
\end{eqnarray}

By analyzing the first condition, we obtain following equation of the energy eigenvalue $E_{n,l}$:
\begin{eqnarray}
E^2_{n,l}&=&M^2+k^2+2\,M^2\,\omega^2\,b_1\,b_2+2\,M\omega\,b_1+2\,\Omega\left(n+1+\sqrt{l^2+M^2\,\omega^2\,b^2_{2}+\eta^2_{c}}\right)\nonumber\\
&&+2\,\eta_L\,\eta_c-\frac{\tilde{b}^2}{4\,\Omega^2}.
\label{41}
\end{eqnarray}
Note that Eq. (\ref{41}) is not the general expression of the relativistic energy eigenvalues of a generalized KG-oscillator. 

The corresponding wave-functions are given by
\begin{equation}
\psi_{n,l} (x)=x^{\sqrt{l^2+M^2\,\omega^2\,b^2_{2}+\eta^2_{c}}}\,e^{-\frac{1}{2}\,\left[x+\frac{\tilde{b}}{\Omega^{\frac{3}{2}}}\right]\,x}\,H (x).
\label{42}
\end{equation}
Note that for $\eta_L \rightarrow 0$ and $b_2 \rightarrow 0$, the energy eigenvalues expression (\ref{41}) and the corresponding radial wave-function (\ref{42}) reduces to the result found in Ref. \cite{b10}. 

Now, we evaluate the individual energy levels and eigenfunctions one by one as in \cite{ERFM,a29,a30,a33,a34}. For example, $n=1$, we have $\Theta=2$ and $c_2=0$ which implies
\begin{eqnarray}
&&\Rightarrow \frac{2}{\beta+\tilde{\theta}}\,d_0=\left(\frac{\eta}{1+2\,\tilde{j}}+\frac{\tilde{\theta}}{2}\right)\,d_0\nonumber\\
&&\Rightarrow \Omega^3_{1,l}-\left [\frac{a^2}{2\,\left(1+2\,\tilde{j}\right)} \right]\,\Omega^2_{1,l}-a\,\tilde{b}\,\left (\frac{1+\tilde{j}}{1+2\,\tilde{j}} \right)\,\Omega_{1,l}-\frac{\tilde{b}^2}{8}\,(3+2\,\tilde{j})=0\quad
\label{43}
\end{eqnarray}
a constraint on the potential parameter $\Omega_{1,l}$. 

The allowed values of the angular frequency of the oscillator for the radial mode $n=1$ is 
\begin{equation}
\omega_{1,l}=\frac{1}{M\,b_1}\,\sqrt{\Omega^2_{1,l}-\eta^2_{L}}.
\label{frequency}
\end{equation}

Therefore, the ground state energy level for the radial mode $n=1$ is given by
\begin{eqnarray}
E^2_{1,l}&=&M^2+k^2+2\,M\,\omega_{1,l}\,b_1\,(M\,\omega_{1,l}\,b_2+1)+2\,\Omega_{1,l}\,\left(2+\sqrt{l^2+M^2\,\omega^2_{1,l}\,b^2_{2}+\eta^2_{c}}\right)+2\,\eta_L\,\eta_c\nonumber\\
&&-\frac{\left(\frac{\alpha\,B_0\,\lambda\,\kappa}{2}+2\,M\,\eta_L \right)^2}{4\,\Omega^2_{1,l}}.
\label{44}
\end{eqnarray}
And the ground state eigenfunction is 
\begin{equation}
\psi_{1,l}(x)=x^{\sqrt{l^2+M^2\,\omega^2_{1,l}\,b^2_{2}+\eta^2_{c}}}\,e^{-\frac{1}{2}\,\left[x+\frac{\left(\frac{\alpha\,B_0\,\lambda\,\kappa}{2}+2\,M\,\eta_L \right)}{\Omega^{\frac{3}{2}}_{1,l}}\right]\,x}\,(1+d_1\,x).
\label{45}
\end{equation}
where we have chosen $d_0=1$ and
\begin{equation}
d_1=\frac{1}{\sqrt{\Omega_{1,l}}}\,\left [\frac{M\,\eta_c}{\left (\frac{1}{2}+\sqrt{l^2+M^2\,\omega^2_{1,l}\,b^2_{2}+\eta^2_{c}}\right)}+\frac{\left(\frac{\alpha\,B_0\,\lambda\,\kappa}{4}+M\,\eta_L \right)}{\Omega_{1,l}} \right].
\label{46}
\end{equation}
The lowest energy state Eq. (\ref{42}) plus the ground state wave-function Eqs. (\ref{43})--(\ref{44}) with the restriction (\ref{frequency}) that gives the possible values of the angular frequency of the oscillator is defined for the radial mode $n=1$. 

\section{Conclusions}

The Klein-Gordon oscillator under the effects of the Lorentz symmetry violation was studied in Ref. \cite{b10}. Inspired by this work, We have analysed the behaviour of the generalized Klein-Gordon oscillator by choosing a function $f(r)=b_1\,r+\frac{b_2}{r}$ in the equation under the effects of a linear central potential induced by Lorentz symmetry violation. Furthermore, we have introduced a Cornell-type scalar potential and analyzed the behaviour of the relativistic quantum oscillator. In {\it section 2}, we have chosen the function $f(r)=b_1\,r+\frac{b_2}{r}$ and derived the radial wave equation. For a suitable wave function, Heun's biconfluent differential equation is derived from this radial wave equation. Substituting the power series in the biconfluent differential equation and finally truncating it, the non-compact expression of the energy eigenvalues Eq. (\ref{22}) and the radial wave function (\ref{23}) is obtained. By imposing the truncating condition $d_{n+1}=0$, the possible values of the oscillator frequency Eq. (\ref{24}), the energy level Eq. (\ref{26}), and the radial wave function Eqs. (\ref{27})--(\ref{28}) associated with the lowest state of the quantum system defined by $n=1$ is obtained. We noted that for $b_2 \rightarrow 0$ in the function $f (r)$, the energy eigenvalue expression (\ref{22}) is very similar to the result obtained in Ref. \cite{b10} (see Eq. (19) in Ref. \cite{b10}). 

In {\it section 3}, we have introduced a Cornell-type scalar potential $S(r)$ in the equation by modifying the mass term under the effects of Lorentz symmetry violation. Following a similar procedure, we have obtained the ground state energy level (\ref{44}) and the lowest state wave function Eqs. (\ref{45})--(\ref{46}) with the restriction (\ref{frequency}) on the angular frequency of the oscillator. Here also, for $b_2 \rightarrow 0$ and $\eta_L \rightarrow 0$, the energy eigenvalues expression (\ref{42}) and the wave-function (\ref{43}) is similar to the result obtained in Ref. \cite{b10}. Thus the linear potential term $\eta_L\,r $ present in the scalar potential $S(r)$ and the extra Coulomb-like term $\frac{b_2}{r}$ in the function $f(r)$ modified the energy spectrum in comparison to the previous result. In both cases, we have seen that for the function $f(r)=b_1\,r+\frac{b_2}{r}$, and the Lorentz symmetry violation parameters $(\alpha, \lambda, B_0, \kappa)$ (in {\it section 2}), and the Cornell-type scalar potential considered in {\it section 3} modified the energy levels and the wave-function associated with each radial mode.

\end{document}